\documentclass[fleqn,usenatbib]{mnras}

\usepackage{newtxtext,newtxmath}
\usepackage{lscape}
\usepackage[capitalise]{cleveref}
\usepackage{threeparttable}
\usepackage{caption3}
\usepackage[normalem]{ulem}
\usepackage{array}

\usepackage[T1]{fontenc}
\usepackage{ae,aecompl}

\usepackage{graphicx}	
\usepackage{amsmath}	
\usepackage{amssymb}	

\newcommand{\msun}{\,M$_\odot$}
\newcommand{\kms}{\,km\,s$^{-1}$}

\newcommand{\cow}{\mbox{AT\,2018cow}}
\newcommand{\host}{\mbox{CGCG\,137-068}}
\newcommand{\ha}{H$\alpha$}
\newcommand{\hii}{{\sc Hii}}
\newcommand{\hi}{{\sc Hi}}
\newcolumntype{H}{>{\setbox0=\hbox\bgroup}c<{\egroup}@{}} 

\title{Studying the environment of \cow{} with MUSE}

\author[J. D. Lyman et al.]{
J.~D.~Lyman,$^{1}$\thanks{E-mail: j.d.lyman@warwick.ac.uk (JDL)},
L. Galbany,$^{2}$
S.~F.~S\'anchez,$^{3}$
J.~P.~Anderson,$^{4}$
H.~Kuncarayakti$^{5,6}$
\vspace{0.2cm}\\
$^{1}$Department of Physics, University of Warwick, Coventry CV4 7AL, UK\\
$^{2}$Departamento de F\'isica Te\'orica y del Cosmos, Universidad de Granada, E-18071 Granada, Spain\\
$^{3}$Instituto de Astronom\'ia, Universidad Nacional Aut\'onoma de  M\'exico Circuito     Exterior, Ciudad Universitaria, \\Ciudad de M\'exico 04510,  Mexico\\
$^{4}$European Southern Observatory, Alonso de C\'ordova 3107, Vitacura, Casilla 190001, Santiago, Chile\\
$^{5}$Tuorla Observatory, Department of Physics and Astronomy, FI-20014 University of Turku, Finland\\
$^{6}$Finnish Centre for Astronomy with ESO (FINCA), FI-20014 University of Turku, Finland\\
}

\date{Accepted XXX. Received YYY; in original form ZZZ}

\pubyear{2020}

\begin{document}
\label{firstpage}
\pagerange{\pageref{firstpage}--\pageref{lastpage}}
\maketitle

\begin{abstract}
\cow{} was the nearest and best studied example of a new breed of extra-galactic, luminous and rapidly-evolving transient. Both the progenitor systems and explosion mechanisms of these rapid transients remain a mystery -- the energetics, spectral signatures, and timescales make them challenging to interpret in established classes of supernovae and tidal disruption events. The rich, multi-wavelength data-set of \cow{} has still left several interpretations viable to explain the nature of this event. In this paper we analyse integral-field spectroscopic data of the host galaxy, \host{}, to compare environmental constraints with leading progenitor models. We find the explosion site of \cow{} to be very typical of core-collapse supernovae (known to form from stars with $M_\textrm{ZAMS}\sim8-25$\,\msun{}), and infer a young stellar population age at the explosion site of few\,$\times 10$\,Myr, at slightly sub-solar metallicity. When comparing to expectations for exotic intermediate-mass black hole (IMBH) tidal disruption events, we find no evidence for a potential host system of the IMBH. In particular, there are no abrupt changes in metallicity or kinematics in the vicinity of the explosion site, arguing against the presence of a distinct host system. The proximity of \cow{} to strong star-formation in the host galaxy makes us favour a massive stellar progenitor for this event.

\end{abstract}

\begin{keywords}
(stars:) supernovae: individual: \cow{} -- galaxies: individual: \host{} -- (stars:) supernovae: general -- stars: massive
\end{keywords}



\section{Introduction}

Newly discovered rapidly evolving, luminous extra-galactic transients have proven a challenge to explain \citep[e.g.][]{drout14, tanaka16, pursiainen18}. The discovery of such transients are a challenge for our existing understanding. Although the samples share some characteristics, studies are in their infancy and inherent diversity amongst the samples is likely to exist, this has led to such transients being named variously as FBOTs (fast and blue optical transients), FELTs (fast-evolving luminous transients) and RETs (rapidly-evolving transients), among others. Their discovery in significant numbers has been brought on by advances in cadence and depths of sky surveys in recent times. Rising typical in only a few days to luminosities exceeding typical supernovae (SNe; which typically rise on timescales of weeks, e.g. \citealt{taddia15}), and decaying similarly rapidly poses a problem for both progenitor and explosion models, and our understanding of the final fates of stars.

Samples of such rapidly-evolving events now exist, although their nature often precludes intensive study since they must be ear-marked as interesting in real time to observe their behaviour before the onset of rapid decay -- given the typical distances of the majority of events, they quickly fade below feasible observational limits. This changed with the discovery of \cow{}, discovered by the Asteroid Terrestrial-impact  Last  Alert  System (ATLAS; \citealt{tonry18}) at a distance of only $\sim61$\,Mpc, in the spiral galaxy \host{} \citep{prentice18}.

An intensive, multi-wavelength campaign ensued for this source. As well as copious UV, optical and infra-red spectra and photometry \citep[e.g.][]{prentice18, perley19, kuin19}, the source was also detected at X-ray \citep{rivera18, margutti19}, millimeter \citep{ho19} and Radio \citep{bietenholz20} wavelengths. 
\cow{} displayed characteristics that matched some of the criteria for several transient models, but no single model was able to match its full behaviour.
For example, the optical spectra appeared reminiscent of stripped-envelope core-collapse SNe (CCSNe) -- those CCSNe with absent or tenuous hydrogen signatures -- but at consistently higher temperatures and photospheric velocities than expected. Further, the tightly-constrained rapid rise to a high peak luminosity of $M_g\sim-20.4$\,mag in only $\sim$2.5\,days \citep{perley19}, and subsequent power-law decay of the light-curve, do not fit well in the radioactively-powered paradigm of CCSNe. 

Additional energy sources have been suggested as significant contributors to the luminosity of \cow{}. Firstly, circumstellar interaction between the SN ejecta and the surrounding medium \citep[see][]{chevalier17} would be an attractive additional energy-source, especially given the presence of interaction signatures in the spectra of \cow{} and similarities in behaviour to interacting classes such as SNe~Ibn \citep{fox19, karamehmetoglu19}.

Central engine activity, such as in the collapsar model of CCSNe \citep{woosley93} is invoked as an additional source of energy for gamma-ray-burst SNe (GRB-SNe) -- i.e. those SNe that are associated with a long-duration GRB. The accretion of stellar material onto a central black-hole (BH), formed during the SN collapse, provides additional energy injection into the SN ejecta in a strongly bi-polar orientation, eventually breaking the stellar surface to produce collimated relativistic jets. This results in the typically larger explosion energies and luminosities and temperatures of GRB-SNe cf. standard CCSNe \citep{lyman16, modjaz16, kann19, taddia19}. The geometric alignment of these bi-polar jets and the Earth may be the single greatest factor in determining whether a GRB is detected associated with a collapsar SN. Nevertheless, very late time radio data can give clues as to the presence of relativistic material largely independent of the jet direction \citep[e.g.][]{soderberg10}. Recent results for \cow{} place limits on any relativistic outflow powered by a central-engine, which, if it was produced, must have been only short-lived \citep{bietenholz20}.

In addition to a massive stellar origin, \cow{} shares a number of similarities with tidal-disruption events (TDEs). In particular the power-law decay of the light-curve is a behaviour seen in typical TDEs in the nuclei of galaxies \citep[e.g.][]{arcavi14, holoien16}, where the central super-massive BH is responsible for the disruption of a main sequence star. The timescales for nuclear TDEs, however, are typically weeks--months. \citet{perley19} applied scaled relations to \cow{} in order to obtain a potential progenitor scenario involving an intermediate mass BH (IMBH; $\sim$10$^4$\,\msun{}) and main sequence star, whereas \citet{kuin19} proposed a more massive BH ($\sim$10$^6$\,\msun{}) disrupting a low-mass white dwarf as an explanation.

The rapidly-evolving nature of transients similar to and including \cow{} gives a small, finite window for their direct study. However, a wealth of literature exists on probing explosive transient progenitors through analysis of their host galaxies and the explosion environments within them. Such studies are benefiting from advents in integral-field spectroscopy (IFS), allowing the environments to be probed in ever greater details alongside their host galaxies \citep[e.g.][]{lyman18, galbany18, kuncarayakti18}.

The host galaxy \host{} has been studied at radio wavelengths, in \hi{} 21cm emission and continuum, by \citet{roychowdhury19} and \citet{michalowski19}. Both studies find a ring-like morphology of the atomic \hi{} gas distribution in \host{}, but infer a different origin for this structure (interaction with an external galaxy, or due to resonance from the bar, respecively for the two studies.) Given the same information, the studies make conflicting interpretations on the nature of the progenitor. \citet{roychowdhury19} suggest the detection of \hi{} indicates the presence of compact star forming regions, giving a viable massive stellar progenitor route for \cow{}. \citet{michalowski19} conversely use an argument that the \hi{} distribution is not as asymmetric or as concentrated at the explosion site as seen in a handful of GRB-SNe hosts to argue that \cow{} may not have been formed by a massive star. Using ALMA data, \citet{matsui19} found the explosion site of \cow{} to share similarities with Type I CCSNe and the host galaxy \host{} to be a typical star-forming dwarf in the local Universe.

Here we present Very Large Telescope/Multi Unit Spectroscopic Explorer (VLT/MUSE) integral-field spectrograph (IFS) data on \host{} and the explosion site of \cow{} within, in order to investigate the nature of the progenitor system.
Throughout we assume a distance of 60.9 Mpc to \host{} based on a redshift of 0.01406 \citep{aguado19} using a $\Lambda$CDM cosmology with $H_0 = 70$\kms{}\,Mpc$^{-1}$ and $\Omega_M = 0.3$. Distance dependant quantities do not include any uncertainty on the distance to \host{}.

\section{Observations and Data Reduction}
\label{sect:obs}

\begin{figure*}
	\includegraphics[width=\textwidth]{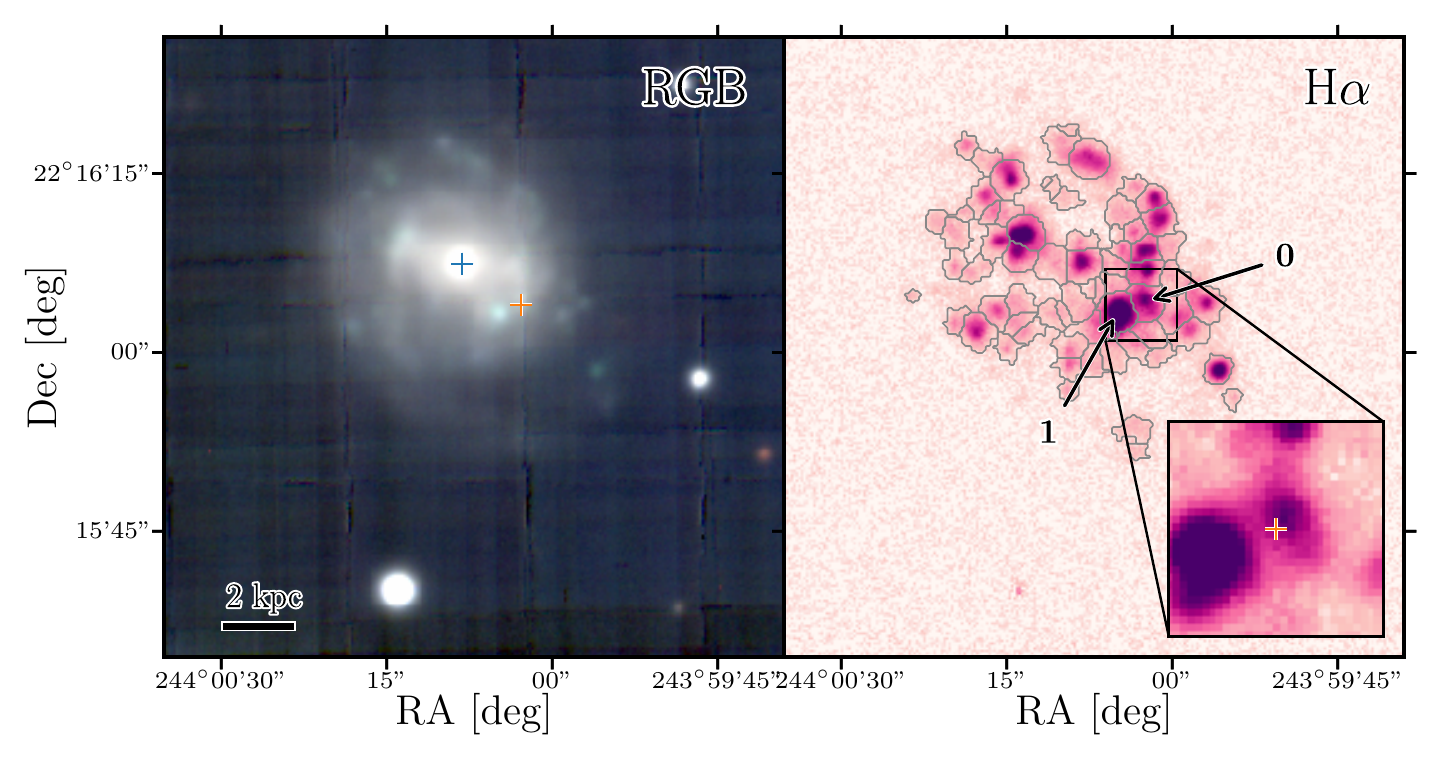}
    \caption{A false colour image \textit{(left)} and a continuum-subtracted \ha{} map \textit{(right)} of \host{} reconstructed from the MUSE data cube. Overlaid on the \ha{} map are contour lines indicating the bins used for \hii{} region analysis -- their creation is described in \cref{sect:analysis}. The blue and orange crosses in the left panel denote the adopted locations of the nucleus of \host{} and the explosion site of \cow{}, respectively. The explosion site is also indicated by the orange cross in the inset right panel.
    Within the inset square, which is $\sim$1.7\,kpc on a side, are shown labels for two \hii{} regions: Regions 0 and 1, which we refer to throughout the analysis. These are the nearest underlying \hii{} region to the explosion site, and the (also nearby) brightest \hii{} region in \host{}, respectively.}
    \label{fig:museimage}
\end{figure*}

We observed \host{}, the host galaxy of \cow{} with MUSE \citep{bacon10} mounted on UT4 of the VLT in Paranal on 2019-05-22, almost 1 year after the discovery of \cow{}, observations were taken as part of the All-weather MUse Supernova Integral -field of Nearby Galaxies programme (AMUSING; \citealt{galbany16}). MUSE is an integral-field unit (IFU) instrument offering seeing-limited spatially resolved spectroscopy over a $\sim1$\,arcmin field of view, large enough to cover the full extent of \host{}. The total exposure was 2805s split over 4 exposures, which were rotated in 90 degrees steps and offset slightly to combat detector artefacts in the final data cube. The exposures were reduced and combined within the {\sc EsoReflex} environment from ESO \citep{freudling13}, using MUSE pipeline version 2.6.2 and sky-residuals were removed using blank regions of the field-of-view and ZAP version 2.1 using default parameters \citep{soto16}.
We determined the PSF FWHM to be 0.86\,arcsec, as measured from isolated stars in the flatted white-light image of the cube. A reconstructed colour image of the cube is shown in \cref{fig:museimage}.

As a final calibration step, we applied a flux scaling factor of 0.35 to the entire data cube. This value was found from the photometric re-calibration analysis as part of AMUSING Data Release 1 from Galbany et al. (in prep), which matches MUSE spectra to archival photometric data from the Sloan Digital Sky Survey \citep[SDSS][]{fukugita96}, PanSTARRS \citep{tonry12} and the Dark Energy Survey \citep{desdr1} for the AMUSING$++$ compilation of MUSE-observed galaxies \citep{lopezcoba20}. We expect an uncertainty on this factor comparable to the photometric accuracy of the matching (few hundredths), which is not a significant source of uncertainty for our results.

\section{Data Analysis}
\label{sect:analysis}

For our MUSE data analysis, we used {\sc ifuanal}\footnote{\url{https://github.com/Lyalpha/ifuanal}} \citep{lyman18}. This package incorporates spectral pixel (spaxel) binning algorithms and fits stellar continuua (using {\sc starlight}, \citealt{cidfernandes05}) and emission-lines in these spaxel bins to discern spatially-resolved properties of galaxies in IFU data. The analysis method used is detailed further in \citet{levan17, lyman18} and the documentation,\footnote{\url{https://ifuanal.readthedocs.io/en/latest/}} and mirrors that done on other MUSE host galaxies \citep[e.g.][]{galbany16}. Briefly, the reduced cube was dereddened by $E(B-V) = 0.078$\,mag \citep{schlafly11} using an $R = 3.1$ \citet{cardelli89} extinction law and deredshifted. Large circular masks were placed over the isolated foreground stars in the field to eliminate them from our analysis. The remaining spaxels were then binned to create distinct regions of the host which are analysed using a summed spectrum.
In order to facilitate our primary emission-line analysis, we utilised a binning algorithm designed to segment \hii{} regions \citep[expanded from][]{sanchez12}. A narrow-band \ha{} map (smoothed with a 0.5\,pixel Gaussian filter to remove noise spikes) was constructed from the cube to determine seeds for spaxel bins as peaks in this image. These bin seeds were grown to pixels satisfying the following criteria: within 0.5\,kpc of the seed, at least 8\,per~cent of the seed pixel's flux and at least 2$\sigma$ above the background level of the \ha{} map. Where pixels lay in overlapping regions between two seeds, their assignment was to the nearest bin seed, with the distance weighted by the flux of each initial bin (prior to this assignment) to the one third power. This produced 66 bins. In addition, three circular aperture custom bins were added:
\begin{enumerate}
    \item The explosion site of \cow{} -- 1.74\,arcsec radius centred on explosion site, to give a 1\,kpc diameter aperture for comparison to literature samples (see \cref{sect:other_environments}). 
    \item The nucleus of \host{} -- 1\,arcsecond radius centred on the nucleus, to simulate an SDSS fibre.
    \item The integrated light of \host{} -- 12\,arcsecond radius centred on the nucleus.
\end{enumerate}
The location of the explosion site bin was determined from a deep, late-time William Herschel Telescope (WHT) $r$-band image, first presented in \citet{perley19}. An affine transformation between the WHT image and a white-light image of the MUSE cube was performed with 8 sources in common using {\sc spalipy}\footnote{\url{https://github.com/Lyalpha/spalipy}}. Root mean square centroid residuals were at the 0.3 pixel level, thus not contributing a significant source of uncertainty for our explosion site analyses. The integrated bin radius was selected from inspection of the flattened white-light image of the cube as the extent of detected signal from \host{}.

Each bin (including the three custom bins) was fit for the continuum using a set of \citet[][2016 update]{bc03} simple stellar population models from the MILES spectral library \citep{sanchezblazquez06} with a \citet{chabrier03} stellar initial mass function (IMF) from 0.1-100\msun{}. The components for the base models comprised 16 ages from 1\,Myr to 13\,Gyr for each of 4 metallicities (Z = 0.004, 0.008, 0.02, 0.05). Emissions lines were fit using a series of Gaussians to obtain fluxes and line-of-sight velocities. Emission line fluxes were corrected based on the Balmer decrement assuming an expected ratio of $F_{\textrm{H}\alpha} / F_{\textrm{H}\beta} = 2.86$ \citep[assuming Case B recombination,][]{osterbrock06}, which provided our estimated gas-phase extinction, $E(B-V)_\textrm{gas}$.

\section{\host{} Results}
\label{sect:results}

\subsection{Stellar Continuum}
\label{sect:stellar_cont}

\begin{figure}
	\includegraphics[width=\linewidth]{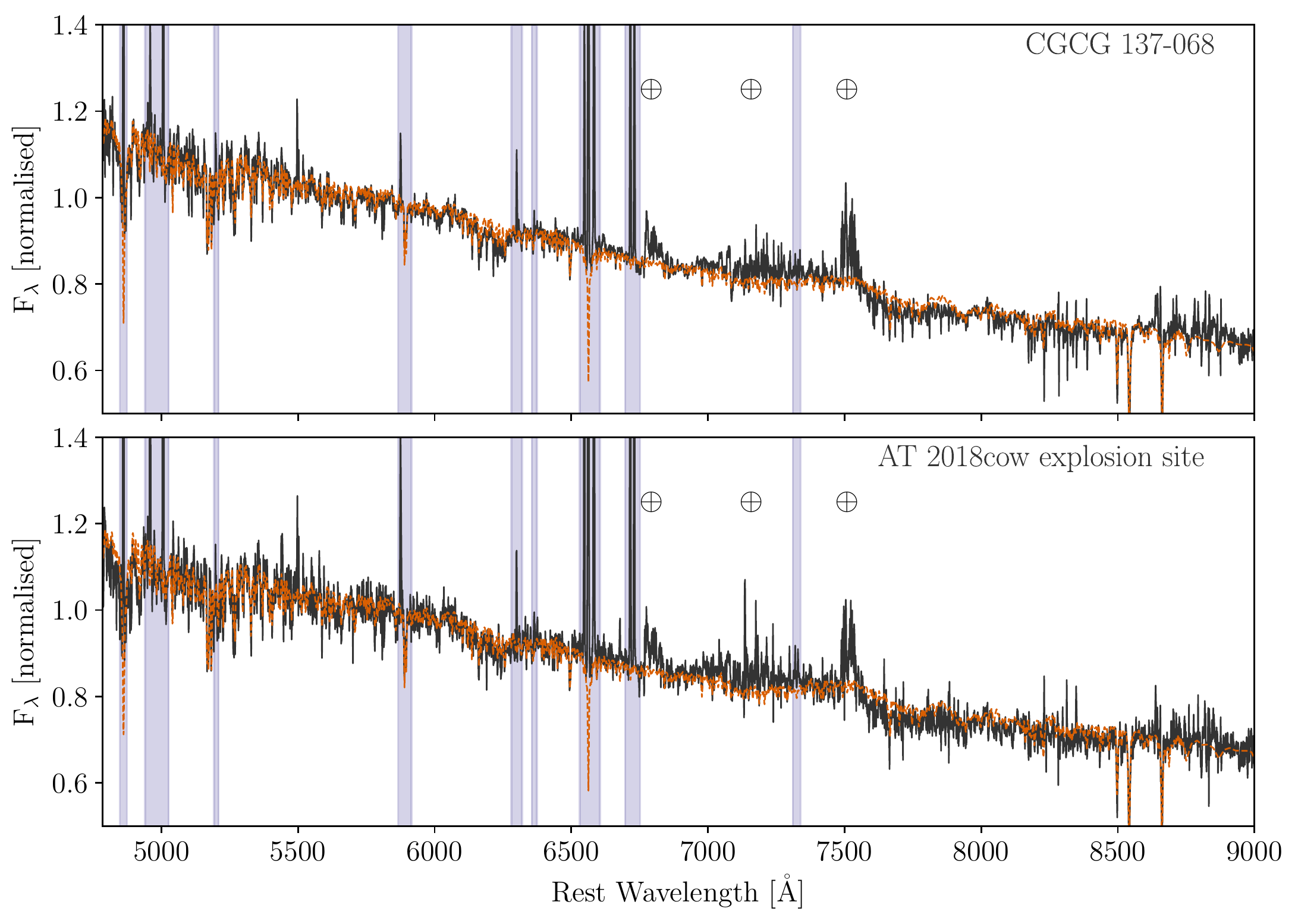}
    \caption{Top: Stellar population fit from {\sc starlight} for \host{} showing extracted MUSE spectrum (black) with model continuum fit (orange, dashed). Shaded regions were masked from the fitting procedure. 
    Bottom: As above but for the explosion site of \cow{}. This region appears to closely resemble the integrated spectrum, making the explosion site typical of the host galaxy.
    Spectra are normalised to the flux in the range 5590--5680\AA{}. 
    }
    \label{fig:sfh}
\end{figure}

Our continuum fits for \host{} and the explosion site of \cow{} are shown in \cref{fig:sfh}. Based on the best fitting star-formation history used to produce these fits, the location of \cow{} does not appear significantly different from the overall host galaxy, containing a similar fraction of young stars, at a few-percent by mass, and being dominated by a solar--sub-solar stellar population of 10 Gyrs. 
We restrict our interpretation to these simple statements since the wavelength range of MUSE is limited -- in particular not extending to blue wavelengths -- meaning we are subject to significant uncertainty and degeneracy in our fitting, particularly when investigating the young stellar populations. The primary aim of the stellar-continuum fitting is to provide a good model continuum to subtract from our spectra to produce pure emission-line spectrum for our subsequent analysis.

Despite the caveats above, we may obtain an estimate of the total galaxy mass as this is less subject to the influence (and degeneracies) of fitting young stellar populations. From fitting the integrated spectrum of \host{}, we obtain a current stellar mass of \mbox{$M_\star = 1.74\substack{+0.07 \\ -0.06}\times10^9$\,\msun{}}. Quoted 1$\sigma$ uncertainties were determined from repeating the fitting on 500 realisations of the integrated spectrum, sampled from its flux and uncertainty, the adopted value is the mean of these 500 values.
A host galaxy mass of \mbox{$M_\star = 1.42\substack{+0.17 \\ -0.29} \times10^9$\,\msun{}} was found from spectral energy distribution (SED) fitting by \citet{perley19}, also assuming a \citet{chabrier03} stellar IMF. The two values are in good agreement ($1.8\sigma$) considering the slightly different distances adopted (our mass becomes \mbox{$M_\star =  1.69\substack{+0.07 \\ -0.06}\times10^9$\,\msun{}} at $D_L = 60$\,Mpc, the value used in \citealt{perley19}), and the different choice of stellar population bases used.

\subsection{In emission}
\label{sect:in_emission}

\begin{figure}
	\includegraphics[width=\columnwidth]{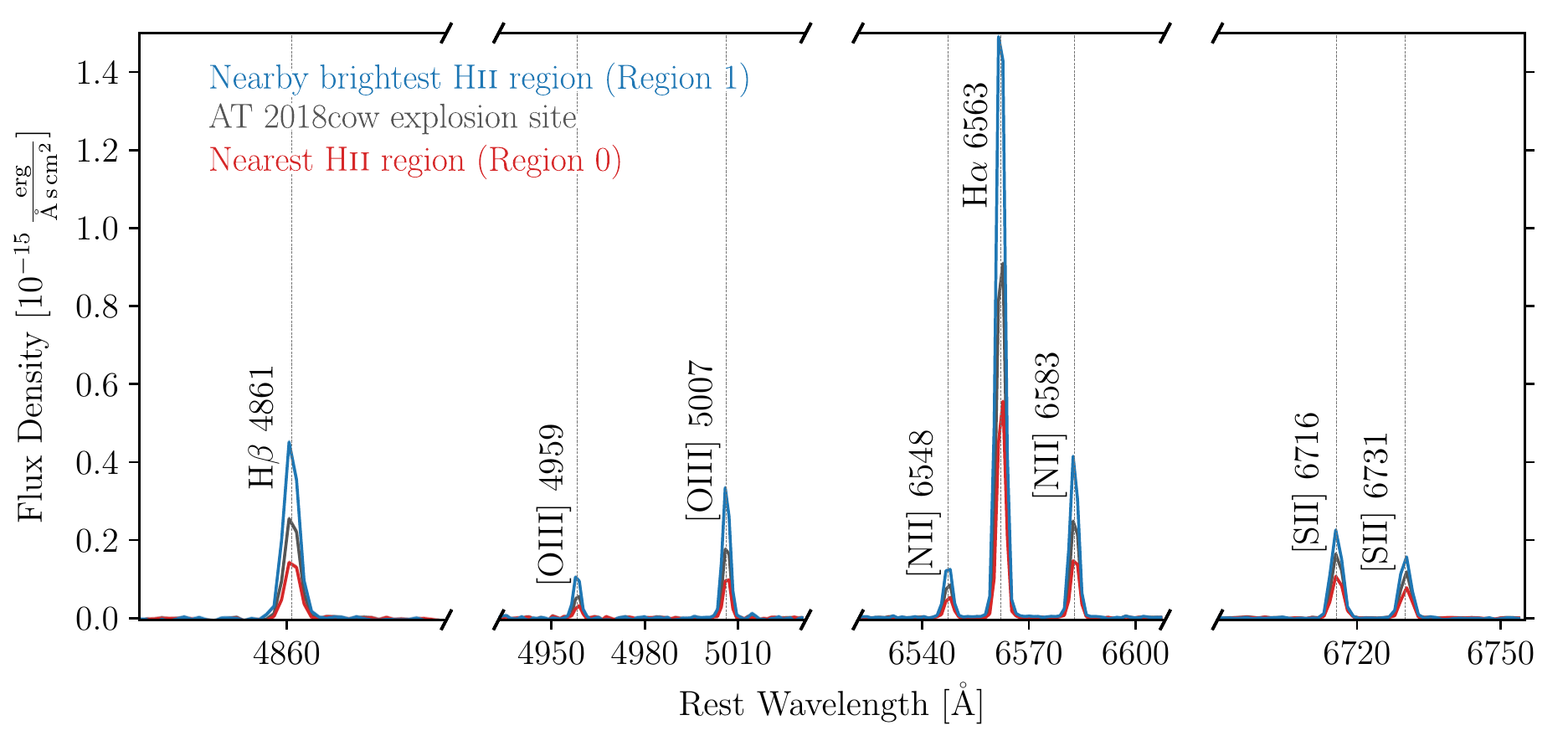}
    \caption{Continuum-subtracted spectra extracted at the regions of interest in the analysis (see \cref{fig:museimage}). Spectra have been cut to highlight the strong-lines of interest. Line fluxes are given in \cref{tab:emission_fluxes}.
    }
    \label{fig:emission_line_fits}
\end{figure}

Emission line ratios from each of our spaxel bins (including the nucleus) lay in the region of normal emission of \hii{} regions \citep{kewley13, sanchez19}, i.e. the ionising radiation is dominated by the contributions of hot, young, massive stars. We thus consider the nebular emission we see in the host as being overwhelmingly due to star-formation and treat it as such. We use the calibration of \citet{dopita16} to determine gas-phase abundances following the recommendation of \citet{kruhler17} for MUSE data. Portions of the continuum-subtracted spectra for our regions of interest are shown in \cref{fig:emission_line_fits} and line fluxes are presented in \cref{tab:emission_fluxes}.

From our host nucleus extracted spectrum, we recover metallicities in agreement with SDSS-based values \citet{matsui19}, namely: $Z = 8.62\pm0.01$ (N2) and $8.72\pm0.03$ (O3N2) dex, using the calibrators of \citealt{pettini04}.

Despite a somewhat regular face-on disk morphology in the continuum, a faint tidal tail is shown in the MUSE colour image extending South West, and the distribution of ongoing star-formation, as traced by \ha{}, is quite irregular, significantly asymmetrical, and weighted in the direction of this tail (\cref{fig:museimage}). This would suggest relatively recent dynamical interaction and/or gas accretion in the history of \host{}. A similar conclusion was drawn by \citet{roychowdhury19} based on the presence of an atomic \hi{} gas ring.
     
The point-like source to the southeast of the explosion site (Region\,1) is confirmed by the MUSE data as a young, intensely star-forming \hii{} region, as predicted by \citet{perley19}. Indeed, this is the most intensely star-forming region of the galaxy with $\log_{10}L_{\textrm{H}\alpha} = 39.52$\,erg\,s$^{-1}$
In our data we discover a further relatively bright \hii{} region almost directly underlying the explosion site (Region\,0), which was obscured in earlier data containing emission from \cow{}, having $\log_{10}L_{\textrm{H}\alpha} = 39.13$\,erg\,s$^{-1}$, about 100 times that of Orion \citep{kennicutt84}. 
The total \ha{} luminosity of \host{}, as determined from the integrated bin, is \mbox{$\log_{10}L_{\textrm{H}\alpha} = 40.37$\,erg\,s$^{-1}$}, making it wholly unremarkable in the local Universe. 

Converting from \ha{} luminosity to SFR using the calibration of \citet{kennicutt98}, we obtain a total SFR$_{\textrm{H}\alpha} = 0.19$\,\msun{}/year, in excellent agreement with the value of $0.22\substack{+0.03 \\ -0.04}$\,\msun{}/year inferred by \citet{perley19} from SED fitting of \host{}. 

\subsection{Explosion site of \cow{}}

The adopted explosion site of \cow{} lies 0.45\,arcsec ($\sim$130\,pc at the distance of \host{}, in projection) from the peak of Region\,0, and 2\,arcsec from Region\,1 ($\sim$570\,pc). Although Region\,1 is the most intensely star-forming region of the galaxy, it does not differ significantly from Region\,0 (\cref{tab:emission_results}), perhaps being slightly younger and metal-rich. Thus the ambiguity in the parent \hii{} region of \cow{} does not affect significantly the environmental constraints on the progenitor. (Any such constraints from the \hii{} regions assume the progenitor was formed recently, and thus coeval with the ongoing star-formation of these regions, see \cref{sect:other_environments}.)

The presence of \hii{} regions almost exactly underlying the explosion site hints towards a causal link between \cow{} and these regions -- the spatial coverage of similarly bright regions as compared to the continuum distribution (\cref{fig:museimage}) is low.

The largely uncertain nature of the progenitor \cow{}, and its spectral similarity to some interacting SNe around maximum light \citep{fox19} merits a search for late-time nebular or interaction emission-line signatures at its explosion site. A manual inspection of the continuum-subtracted spectrum revealed no sources of flux that we could not ascribe to normal \hii{} region emission.

\begin{figure}
	\includegraphics[width=\columnwidth]{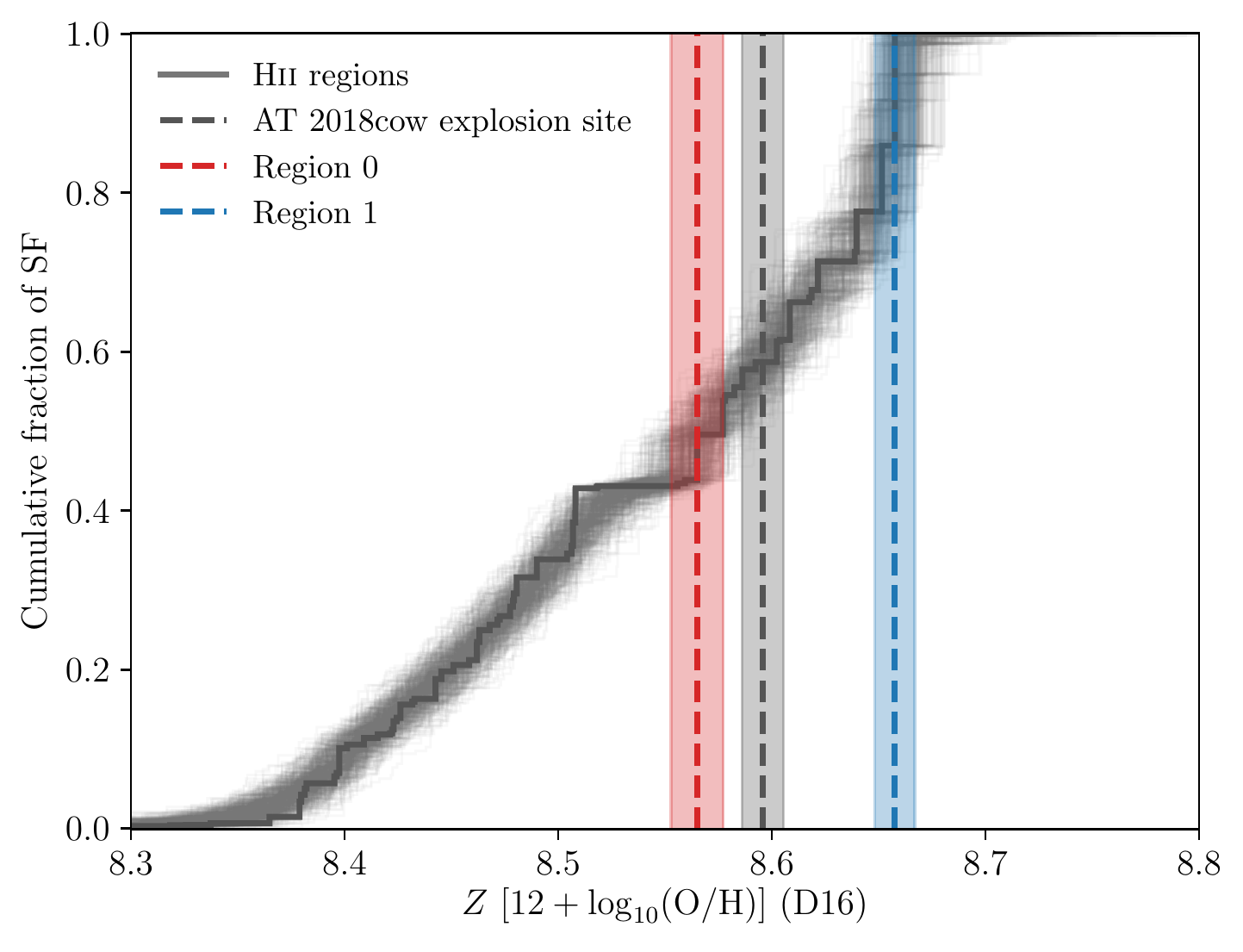}
    \caption{Cumulative distributions showing the distributions of star-formation as a function of metallicity in \host{} using the indicator of \citet{dopita16}. 
    The \hii{} regions contribute a weight to the distribution given by their \ha{} luminosity, making the distribution a cumulative fraction of ongoing star formation (rather than just raw counting of \hii{} regions). 500 realisations are shown assuming Gaussian statistical uncertainties on the metallicity and \ha{} luminosity weighting.}
    \label{fig:cumulative_metallicity}
\end{figure}

We show the locations of these regions in the cumulative distribution of star-formation throughout \host{} in \cref{fig:cumulative_metallicity}. The host galaxy \hii{} regions contribute a weight to the cumulative distribution given by their (Balmer decrement-corrected) \ha{} luminosity. The distributions thus show the fraction of stars being formed at that metallicity or less, giving the $Z$NCR statistic introduced in \citet{lyman18}. Although we see the choice of metallicity indicator has a noticable impact on the distributions, and in particular the position of Region\,1 within the respective distributions, the explosion site of \cow{} consistently has a $Z$NCR $\sim 0.5-0.6$, indicating it is a typical metallicity for stars being formed in \host{}.

\begin{table*}
\begin{threeparttable}
     \caption{
     Strong-emission line results for regions of interest in \host{} (regions defined in \cref{fig:museimage}). Uncertainties quoted are statistical only.}
     \begin{tabular}{lHcccc}
\hline

Location              & $E(B-V)_\textrm{star}$  & $E(B-V)_\textrm{gas}$ & $L$(H$\alpha$)              & EW(H$\alpha$) & $Z$\tnote{a} \\
                      & mag                    & mag                    & [$\log_{10}$ erg s$^{-1}$]  & [\AA{}]       & $12 + \log_{10}(\textrm{O/H})$  \\
\hline  
\cow{} explosion site & 0.09                   & 0.23                   & $39.33\pm0.01$              & $78.7\pm2.1$  & $8.60\pm0.01$ \\
Region\,0             & 0.09                   & 0.26                   & $39.13\pm0.01$              & $65.9\pm1.9$  & $8.57\pm0.01$ \\
Region\,1             & 0.08                   & 0.19                   & $39.52\pm0.01$              & $111.3\pm2.9$ & $8.66\pm0.01$ \\
\host{} (nucleus)     & 0.08                   & 0.27                   & $38.88\pm0.01$              & $22.3\pm0.7$  & $8.65\pm0.02$ \\
\host{} (integrated)  & 0.10                   & 0.19                   & $40.37\pm0.01$              & $33.9\pm0.9$  & $8.53\pm0.01$ \\
     \end{tabular}
\label{tab:emission_results}
 \begin{tablenotes}
 \item [a] Gas-phase abundance measured in the scale of \citet{dopita16}.
 \end{tablenotes}
\end{threeparttable}
\end{table*}

\section{Comparison to other transients environments}
\label{sect:other_environments}

Here we assess our findings on the environment of \cow{} in the context of known and expected environment properties for plausible progenitor explanations.

\subsection{Massive star progenitor}

\begin{figure}
	\includegraphics[width=\columnwidth]{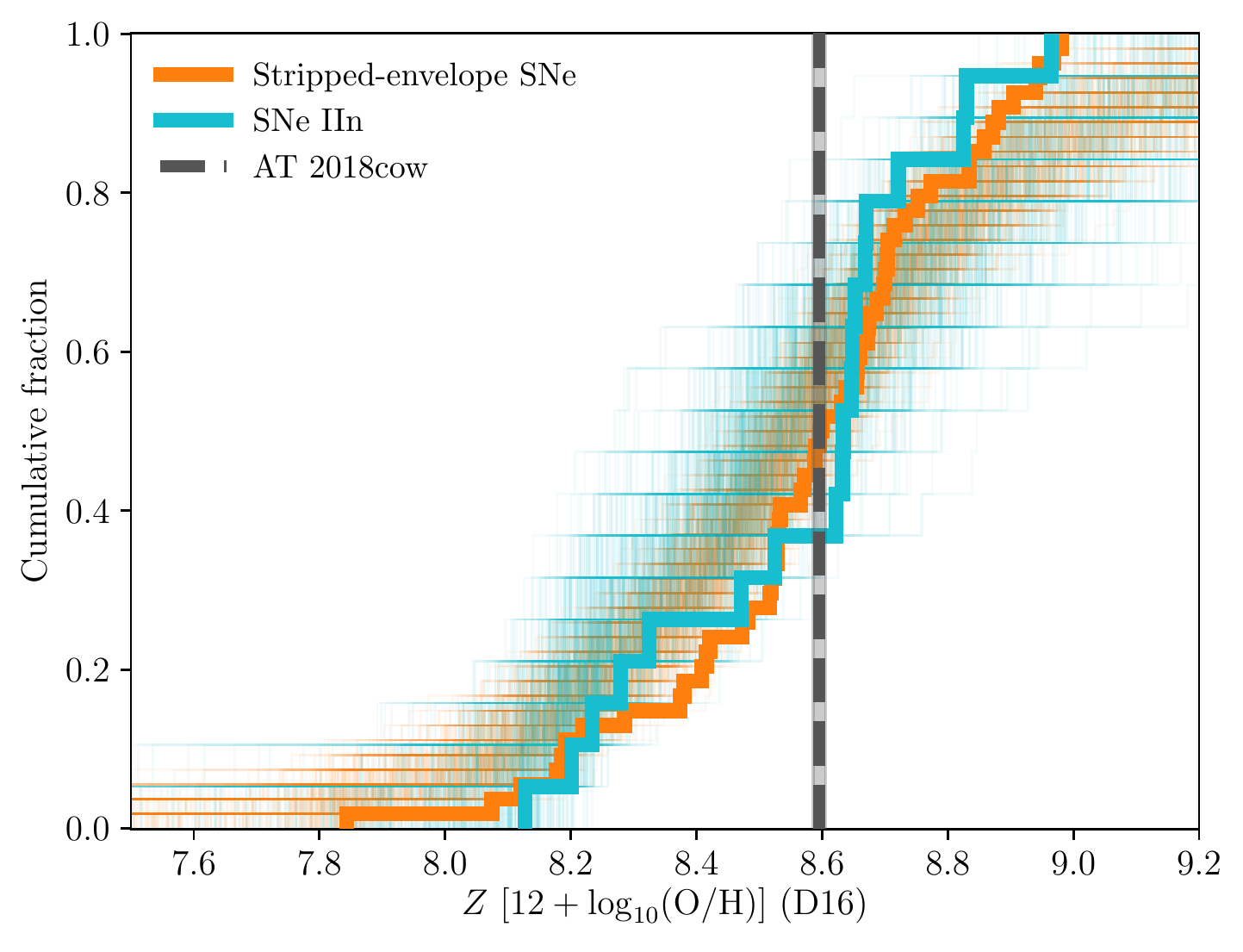}
	\includegraphics[width=\columnwidth]{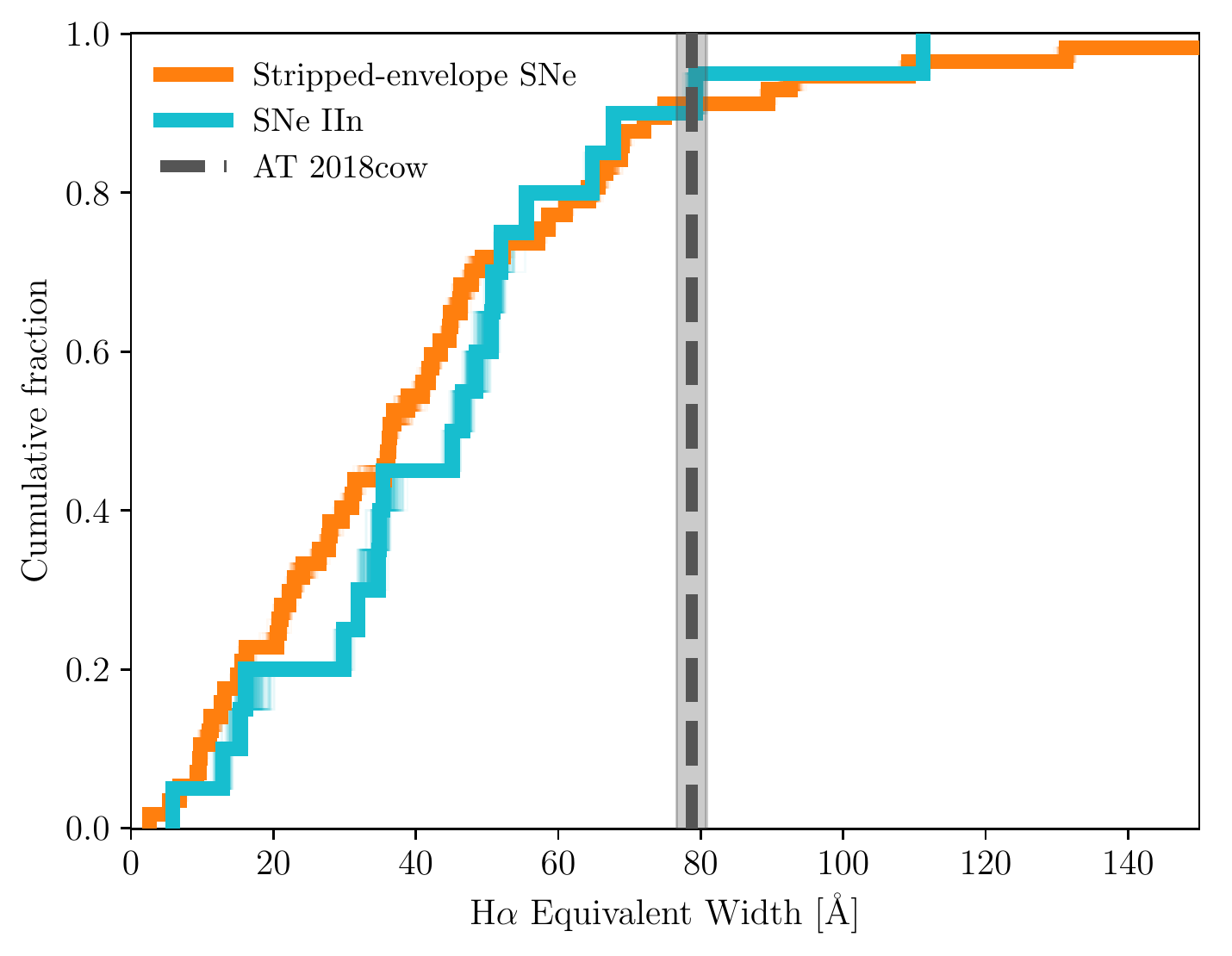}
    \caption{Cumulative distributions of metallicity (top) and \ha{} EW (bottom) for stripped-envelope SN environments in the PISCO sample \citep{galbany18}. Distributions are shown following \cref{fig:cumulative_metallicity}. (The `spreading' of the re-sampled metallicity distributions owes to the relatively large uncertainties compared to the range of values). The explosion site of \cow{} is indicated on each plot. 
    }
    \label{fig:pisco_compare}
\end{figure}

If we take \cow{} to have arisen from a massive stellar explosion, we may expect some similarity to the environments of CCSNe. Indeed, \cow{} being hosted by a later-type galaxy and located in a region of ongoing star-formation, seems to meet the expectations of a young progenitor. For a more quantitative comparison we use the PISCO sample of SN environment properties from \citep{galbany18}. The advantage of using this IFU-based CCSN environment sample is that we are able to compare directly with the sample physical size aperture (1\,kpc). We compare \cow{} to stripped-envelope SN (SESN) sample (i.e. those CCSNe showing absent or tenuous H features) and interacting SNe IIn owing to having the closest similarities in terms of empirical classification with \cow{} \citep[e.g.][]{perley19, fox19} in \cref{fig:pisco_compare}. The explosion site of \cow{} appears almost exactly half way along the SN metallicity distributions, and lies towards the upper end of the \ha EW distribution (indicating a relatively younger environment cf. the median SESN or SN IIn in these comparison samples). However, given the comparison distributions span relative wide-ranges, and we are comparing them with a single object, making any firm inferences on the progenitor of \cow{} from these plots appears to be frivolous. 

It is of interest to compare \cow{} with the environments of SNe~Ibn (i.e. hydrogen-poor SNe with signature of narrow helium lines). \cow{} has been discussed as sharing similarities to SN Ibn both spectroscopically \citep{fox19} and photometrically \citep{karamehmetoglu19}. The progenitors of SN Ibn are currently debated. In the remainder of the paragraph we determine what we may learn about SNe Ibn if we assume \cow{} as being a member of the class -- in the opposing case such arguments would be applicable to only the progenitor of \cow{} and we of course cannot infer any extension to SNe Ibn progenitors. The favoured model of \citet{wang19} for the rapidly-rising and luminous SN~Ibn PS15dpn invokes an interaction $+$ $^{56}$Ni powered CCSN explosion. Given their inferred ejecta mass ($\sim$13\msun{}) is significantly above ejected masses from typical stripped-envelope SNe \citep{lyman16}, the authors suggest a Wolf-Rayet (WR) progenitor for this SN Ibn.
WR stars are amongst the most massive (and thus short-lived -- few Myr) stars. Neither Region\,0 or 1 exhibit the properties of WR regions: i) our \ha{} EW measurements would place the regions at ages of tens of Myr \citep{lyman16, xiao19}, based on comparison with models of stellar populations including multiplicity from the Binary Population and Spectral Synthesis code \citep{eldridge17}; ii) we find no evidence in the continuum-subtracted spectra for emission lines either indicating directly the presence of WR stars (e.g. the red bump, C\,{\sc iii/iv} N\,{\sc iii/iv}) or very young SPs (He\,{\sc i} $\lambda 4922$).\footnote{5$\sigma$ line fluxes were $\sim 0.7-1\times10^{-17}$\,erg\,s$^{-1}$\,cm$^{-2}$ in the regions surrounding \cow{}.} We therefore argue against a WR origin for \cow{}. These arguments also hold for the suggestion of pulsational pair-instability SNe \citep{woosley07} as the progenitors for SNe Ibn \citep{karamehmetoglu19}, since these SNe are expected to be also caused by extremely massive stars (M$_\textrm{ZAMS} \sim$ 100\,\msun{}).
IFS environment studies of samples of SNe Ibn are in progress to address questions about their progenitors, and thus the comparison to \cow{}, more thoroughly.

The maintenance of very high temperatures, its large photospheric velocities \citep{prentice18, perley19} and large X-ray flux \citep{rivera18, kuin19} make \cow{} appear reminiscent of SNe associated with additional energy injection, such as engine-driven events giving rise to long-duration gamma-ray bursts (LGRBs). Such GRB-SNe, and LGRBs in general, preferentially inhabit younger environments compared to other CCSN types \citep[e.g.][]{fruchter06, kelly14, lyman17}. Indeed the host galaxy and specific explosion location of \cow{} is remarkably similar to two of the best studied low-redshift GRB-SNe: GRB~980425/SN~1998bw \citep{kruhler17} and GRB~100316D/SN~2010bh \citep{izzo17}. Both were located coincident with strong 
star-forming regions and close to ($< 1$\,kpc) the most intensely star-forming region of their hosts. There is evidence that LGRBs are subject to a metallicity cut-off, above which their production is suppressed, although the location of this cut-off is debated and may be higher than previously determined, at roughly solar \citep{perley16}. Given our determined metallicity for the explosion site is slightly sub-solar, there does not appear great tension with this cut-off.

The high peak luminosity of \cow{} makes it comparable to super-luminous SNe (SLSNe; e.g. \citealt{galyam12, inserra13}), which are expected to arise from particularly massive stellar explosions with additional energy input thought to arise from magnetar spin-down or circum-stellar interaction. Their host galaxies and environments share a number of similarities with those of LGRBs \citep[e.g.][]{lunnan14, angus16, lyman17}, being generally low-mass and compact. Although no data are available for a direct comparison in our \citet{dopita16} metallicities, a number of studies have noted the strong metal-aversion of SLSN production \citep[e.g.][]{chen17, schulze18}, indicating a threshold of roughly half-solar for their environments. For such a cut-off, our almost solar metallicity determined at the explosion site of \cow{} would place it in tension with this progenitor interpretation, although exceptions exist \citep[e.g. SN~2017egm][]{nicholl17}.

Environmental arguments have been made to suggest \cow{} was not a massive star based solely on the comparison to GRB-SNe environments by observing the form of \hi{} gas distributions of the respective host galaxies \citet{michalowski19}. However, GRB-SNe production is a relatively unknown process that occurs in only a few percent of all CCSNe \citep[e.g.][]{graham16} and indeed could have a causal link or be enhanced with the presence of relatively pristine gas inflows in the local Universe, given the metallicity cut-off seen for LGRB progenitors. Our discovery of strong \hii{} region emission located at the explosion site of \cow{} appears a strong means to link the progenitor with ongoing star-formation, and thus young (massive) stellar populations. This in contrast to inference from observing cold inflows of neutral gas, which provide an even more indirect indicator of current star-formation. We note however that it is very difficult to make conclusive statements on a single object, given almost {\em any} location in a galaxy will have a significant line-of-sight population of old stars.

\subsection{Tidal-disruption event}

Other explanations for \cow{} have been proposed that may not involve a massive star at all. One such scenario is the tidal-disruption of a main-sequence star or white dwarf by an IMBH \citep{perley19, kuin19}.

In the main-sequence TDE scenario \citep{perley19}, a BH mass of $\sim10^{4.3}$\,\msun{} is required. Such an IMBH would be expected to be located within a massive stellar cluster. For the case of a young parent cluster, we do indeed find a star-forming region underlying \cow{}, however the explosion site is significantly offset from the peak of this source in both the white-light image of our MUSE cube and the centroid of the \ha{} emission, where one would expect the massive IMBH to settle towards rapidly. For an older, globular cluster (GC) host, the hosting GC would lie at the upper end of the GC luminosity function \citep{harris96}. A magnitude of the host of M$_V,\sim -9$ to $-11$\,mag follows from both extending the M-$\sigma$ relation \citep[e.g.][]{gultekin09} to IMBH masses, and looking at the luminosities of GCs with signs of IMBHs \citep[e.g.][]{noyola10, lutzgendorf13, feldmeier13}. At the distance of \host{} this would appear as a m$_V \sim 23-25$\,mag source at the explosion site, and should be detectable in high-resolution imaging of the site.

\begin{figure}
	\includegraphics[width=\columnwidth]{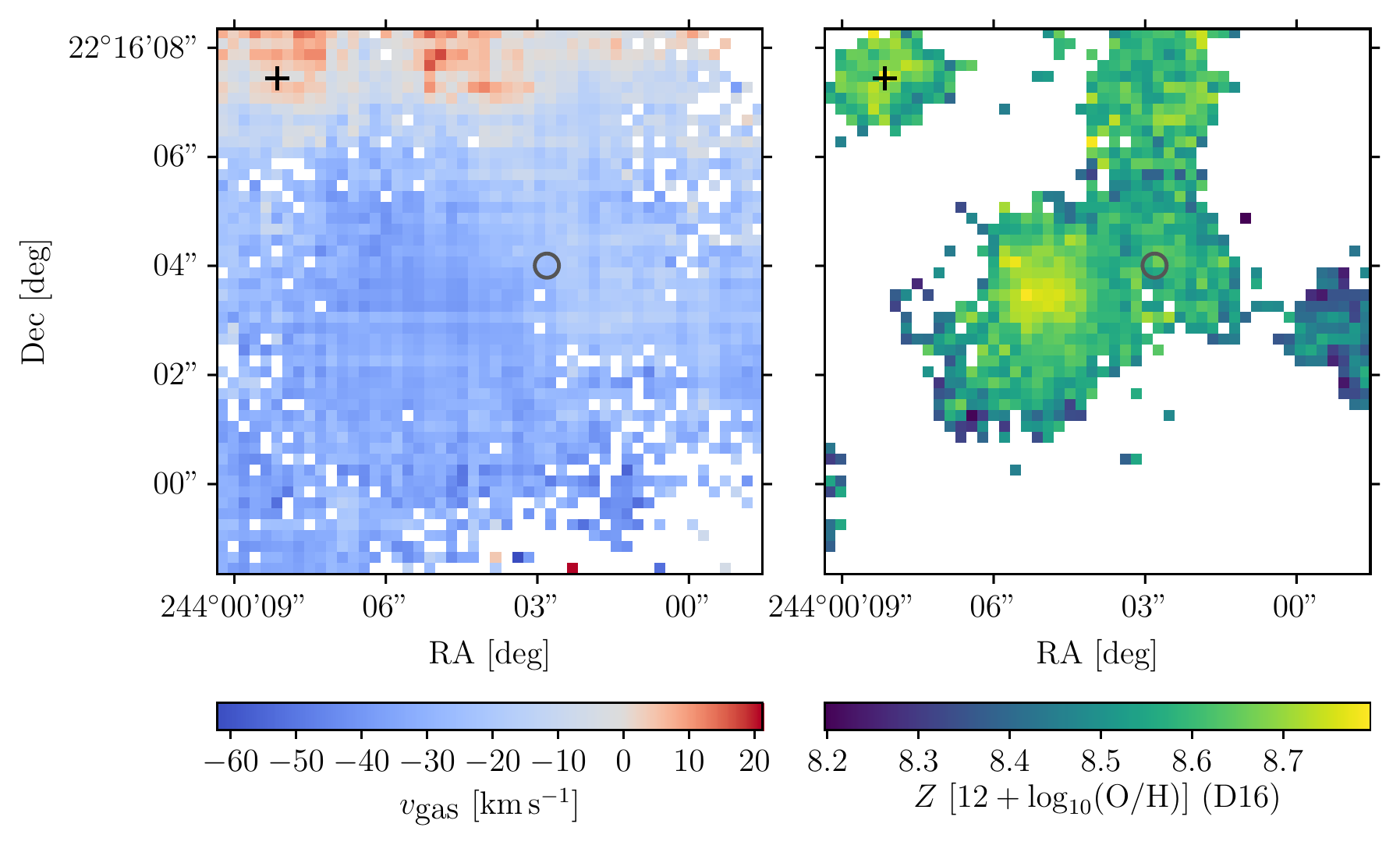}
    \caption{The velocity (left) and metallicity (right) of gas in \host{} surrounding the explosion site of \cow{}. The velocity was determined from the peak of a Gaussian fitted to \ha{} and is shown relative to the adopted redshift of \host{}. White pixels indicate regions where the signal-to-noise ratio of the emission lines prevented the fitting procedure from converging. Given these smoothly evolving maps, we find no evidence of a distinct satellite source underlying \cow{}, which may have appeared as distinct in the velocity and metallicity maps. The explosion site of \cow{} and the nucleus of \host{} are indicated by a circle and cross in each panel, respectively.
    }
    \label{fig:indiv_pix}
\end{figure}

An alternate scenario, involving a more massive BH $\sim10^{5-6}$\,\msun{} and a low-mass white dwarf was presented by \citet{kuin19}. Although white dwarfs may form at a few $\times 10^7$\,yrs after star-formation \citep{portegieszwart07}, i.e. comparable to the age we infer for the underlying \hii{} at the explosion site of \cow{}), these are the most massive white-dwarfs -- low-mass white-dwarfs will form much later. In this case the co-location with star-formation is a coincidence in this scenario. This BH required mass is comparable to that of a massive GC itself, and a dwarf satellite galaxy of \host{} may be a more promising BH-host system. Notwithstanding the lack of spatial coincidence we find between \cow{} and the peak of any underlying source, we searched in our data for existence of a distinct system in the vicinity. In \cref{fig:indiv_pix} we show the results of fitting individual spaxels in a region surrounding the explosion site of \cow{} to look for departures from smoothly varying behaviour in line-of-sight velocity of the gas and stars, and the gas-phase metallicity. We find no evidence for a distinct velocity component in the maps and the line profiles are well modelled by single Gaussians (\cref{fig:emission_line_fits}), arguing against the presence of any satellite galaxy at this location. We also note the metallicity is smoothly varying over this region, whereas a satellite galaxy may be expected to have a different metallicity profile.

\section{Conclusions}

On the balance of evidence presented by our MUSE data, we would favour a young (and therefore likely to be massive) progenitor for \cow{}. We find the transient exploded in close proximity to the most intensely star-forming region of the galaxy and coincident with an underlying \hii{} region. The environment appears typical of other massive star progenitor CCSNe explosion sites based on optical diagnostics, indicating a progenitor age of tens of Myrs. Alternative scenarios, particularly those involving tidal disruptions by IMBHs are less favoured based on a lack of evidence for a massive host system at the site that could plausibly host the IMBH, although deep, high-resolution imaging is required to place meaningful limits on globular clusters. Our deep synthesised optical image of \host{} indicates tidal tails and the \ha{} image indicates asymmetry in the spatial distribution of star-formation, indicative of recent merger history for the galaxy.

A comparison of our environmental results for the very local and fortuitous \cow{} will need to be considered alongside statistical analyses of the environments of more distant similar transient samples in order to elucidate the progenitors and diversity of these fast, luminous transients.

\section*{Acknowledgements}

We thank Dan Perley for kindly providing deep WHT imaging of \cow{} from which to perform relative astrometry. Klaas Wiersema and Elizabeth Stanway are thanked for useful discussions.
JDL acknowledges support from STFC via grant ST/P000495/1.
LG was funded by the European Union's Horizon 2020 research and innovation programme under the Marie Sk\l{}odowska-Curie grant agreement No. 839090. This work has been partially supported by the Spanish grant PGC2018-095317-B-C21 within the European Funds for Regional Development (FEDER).
SFS thanks for the support of a CONACYT grant CB-285080 and FC-2016-01-1916, and funding from the PAPIIT-DGAPA-IN100519 (UNAM) project.
We acknowledge the usage of the HyperLeda database (http://leda.univ-lyon1.fr)
Based on observations collected at the European Organisation for Astronomical Research in the Southern Hemisphere under ESO programme 0103.D-0440(A).
This research made use of Astropy,\footnote{http://www.astropy.org} a community-developed core Python package for Astronomy \citep{astropy13, astropy18}. 



\bibliographystyle{mnras}
\bibliography{references}



\appendix

\section{Emission line fluxes} 

For completeness, individual emission line flux measurements from continuum-subtracted spectra for regions of interest (see \cref{sect:analysis}) are presented in \cref{tab:emission_fluxes}.

\begin{table*}
\begin{threeparttable}
     \caption{Emission line fluxes for regions of interest in \host{}. Fluxes are not corrected for intrinsic extinction, but have been corrected for foreground Galactic extinction (\cref{sect:analysis}). Units are $10^{-15}$\,erg\,s$^{-1}$\,cm$^{-2}$. Uncertainties quoted are statistical only and limits at $3\sigma$.}
     \begin{tabular}{lccccccc}
\hline
Location & H$\beta$ & {\sc [Oiii]} 4959 & {\sc [Oiii]} 5007 & He{\sc i} 5876 & {\sc [Oi]} 6300 & {\sc [Oi]} 6364 & {\sc [Nii]} 6548 \\
\hline
AT~2018cow explosion site    & $ 1.69\pm0.06$ & $ 0.35\pm0.03$ & $ 1.16\pm0.05$ & $ 0.18\pm0.02$ & $ 0.08\pm0.02$ & $<0.04$        & $ 0.44\pm0.01$ \\
Region\,0                    & $ 2.62\pm0.07$ & $ 0.61\pm0.04$ & $ 1.86\pm0.06$ & $ 0.25\pm0.03$ & $ 0.07\pm0.02$ & $<0.04$        & $ 0.63\pm0.02$ \\
Region\,1                    & $ 1.07\pm0.05$ & $ 0.21\pm0.03$ & $ 0.71\pm0.04$ & $ 0.11\pm0.02$ & $ 0.07\pm0.02$ & $<0.03$        & $ 0.29\pm0.01$ \\
CGCG\,137-068 (nucleus)      & $ 0.60\pm0.05$ & $<0.10$        & $ 0.20\pm0.04$ & $<0.07$        & $<0.05$        & $<0.03$        & $ 0.19\pm0.01$ \\
CGCG\,137-068 (integrated)   & $18.53\pm0.80$ & $ 3.37\pm0.48$ & $11.15\pm0.68$ & $ 1.64\pm0.37$ & $ 1.74\pm0.29$ & $ 0.63\pm0.19$ & $ 5.25\pm0.24$ \\
\hline

\hline
Location & \ha{}  & {\sc [Nii]} 6583 & He{\sc i} 6678 & {\sc [Sii]} 6716 & {\sc [Sii]} 6731 & Ar{\sc iii} 7136 & {\sc [Siii]} 9067 \\
\hline  
AT~2018cow explosion site    & $ 4.84\pm0.03$ & $ 1.28\pm0.02$ & $ 0.80\pm0.02$ & $ 0.55\pm0.01$ & $ 0.27\pm0.01$ & $ 0.06\pm0.01$ & $ 0.08\pm0.01$ \\
Region\,0                    & $ 7.49\pm0.04$ & $ 1.88\pm0.03$ & $ 1.00\pm0.02$ & $ 0.70\pm0.02$ & $ 0.45\pm0.02$ & $ 0.08\pm0.01$ & $ 0.14\pm0.02$ \\
Region\,1                    & $ 3.06\pm0.03$ & $ 0.84\pm0.02$ & $ 0.57\pm0.01$ & $ 0.39\pm0.01$ & $ 0.17\pm0.01$ & $ 0.04\pm0.01$ & $ 0.05\pm0.01$ \\
CGCG\,137-068 (nucleus)      & $ 1.71\pm0.03$ & $ 0.55\pm0.02$ & $ 0.32\pm0.01$ & $ 0.22\pm0.01$ & $ 0.07\pm0.01$ & $<0.03$        & $<0.04$        \\
CGCG\,137-068 (integrated)   & $52.99\pm0.43$ & $15.40\pm0.33$ & $11.61\pm0.31$ & $ 7.92\pm0.28$ & $ 2.90\pm0.19$ & $ 0.61\pm0.18$ & $ 0.67\pm0.21$ \\
\hline

\end{tabular}
\label{tab:emission_fluxes}
 \begin{tablenotes}
 \item
 \end{tablenotes}

\end{threeparttable}
\end{table*}

\bsp	
\label{lastpage}
\end{document}